\documentclass[aps,prl,showpacs,twocolumn,superscriptaddress]{revtex4}

\usepackage{amsfonts,amssymb,amsmath}
\usepackage{amsthm,multirow}
\begin{document}

\title{Entropic Version of the Greenberger-Horne-Zeilinger Paradox}

\author{Sadegh \surname{Raeisi}}
\email{sraeisi@uwaterloo.ca}
\affiliation{Institute for Quantum Computing, University of Waterloo, Ontario, Canada}
\affiliation{Centre for Quantum Technologies, National University of Singapore, 3 Science Drive 2, 117543 Singapore, Singapore}

\author{Pawe{\l} \surname{Kurzy\'nski}}
\email{cqtpkk@nus.edu.sg}
\affiliation{Centre for Quantum Technologies, National University of Singapore, 3 Science Drive 2, 117543 Singapore, Singapore}
\affiliation{Faculty of Physics, Adam Mickiewicz University, Umultowska 85, 61-614 Pozna\'{n}, Poland}

\author{Dagomir \surname{Kaszlikowski}}
\email{phykd@nus.edu.sg}
\affiliation{Centre for Quantum Technologies, National University of Singapore, 3 Science Drive 2, 117543 Singapore, Singapore}
\affiliation{Department of Physics, National University of Singapore, 2 Science Drive 3, 117542 Singapore, Singapore}

\date{\today}

\begin{abstract}

Consider four binary $\pm 1$ variables  $A$, $B$, $C$ and $D$ for which classical reasoning implies $ABCD=1$. In this case the knowledge of $A$, $B$, $C$ automatically provides knowledge of $D$ because $D=ABC$. However, the Greenberger-Horne-Zeilinger paradox shows that despite classical prediction one can find quantum states and observables with well defined outcomes for which $D=-ABC$. In this work we formulate an information-theoretic version of this paradox. We show that for a tripartite quantum system one can find a set of four properties for which classical reasoning implies that $D=ABC$, yet quantum theory predicts that one can know everything about $A$, $B$, $C$ and nothing about $D$. 

\end{abstract}

\maketitle

\emph{Introduction.} It was first observed by Bell \cite{Bell} that bipartite quantum systems can violate local realism and later this observation was extended to multipartite systems \cite{Mermin,M2,M3}. It was also shown that for more than two particles it is possible to formulate the Greenberger-Horne-Zeilinger (GHZ) paradox \cite{GHZ}, that is an all-versus-nothing contradiction of local realism, in a sense that it employs deterministic measurement outcomes. Around the same time it was shown that the local realism of bipartite systems can be studied using the notion of Shannon entropy and the bipartite information-theoretic Bell inequalities were proposed \cite{BC}. However, multipartite entropic inequalities have not been constructed up to now. In this work we extend the previous results, provide a tripartite information-theoretic Bell inequality and formulate an entropic version of the GHZ paradox.  

The original GHZ paradox \cite{GHZ} occurs in multipartite quantum systems. In a simplest scenario one considers a system of three qubits in the GHZ state $|GHZ\rangle=\frac{1}{\sqrt{2}}(|000\rangle + |111\rangle)$ and four observables $A=YYX$, $B=YXY$, $C=XYY$ and $D=XXX$, where $X$ and $Y$ are Pauli operators (whose measurement outcomes are $\pm 1$) and we used the simplified notation $XXX \equiv X\otimes X \otimes X$. Any local realistic theory that assigns measurement outcomes to local observables necessarily predicts that the product $ABCD=1$, since in this product each $\pm 1$ observable $X$ and $Y$ occurs twice. However, for the GHZ state all four observables are determined and are $A=-1$, $B=-1$, $C=-1$ and $D=1$ which contradicts local realistic predictions.

Here, we show that similar all-versus-nothing contradiction can be formulated in terms of information. In particular, we define four observables $A$, $B$, $C$ and $D$ for which classical reasoning predicts that if we have full knowledge of $A$, $B$ and $C$, then we automatically have full knowledge of $D$. However, for the GHZ state one has full knowledge of $A$, $B$, $C$ and at the same time no knowledge of $D$ at all. 

We use the previously developed information-theoretic distance approach to nonclassical correlations \cite{distance1,distance2} and propose a new distance-like property that can be applied to binary $\pm 1$ measurements. This property quantifies multipartite correlations in terms of Shannon entropy and can be applied to derive a tripartite inequality whose structure resembles the tripartite Mermin inequality \cite{Mermin}. However, in our case we relate entropies of multipartite measurement outcomes, not multipartite correlation functions like in Ref. \cite{Mermin}. The inequality is satisfied whenever the information-theoretic distance-like property applies to the system (which is true for example in case of classical local realistic systems), but is violated by measurements on quantum systems. Interestingly, our tripartite inequality can be maximally violated within quantum theory, which does not happen in the bipartite case \cite{BC}. Due to this fact, our inequality can be interpreted as an entropic version of the GHZ paradox.

The information-theoretic distance was originally proposed by Zurek \cite{Zurek}. Initially, it was defined using the notion of Kolmogorov entropy, but it can be also defined via Shannon entropy \cite{distance1,distance2} as $d(A,B)=H(A|B)+H(B|A)$, where $A$ and $B$ are random variables, $H(A|B) = H(AB) - H(B)$ is the conditional Shannon entropy, $H(B)$ is the entropy of the variable $B$ and $H(AB)$ is the joint entropy of $A$ and $B$. 
It can be shown that $H(A|B)+H(B|A)$ satisfies all of the requirements of a distance \cite{distance1}. 

The information-theoretic Bell inequalities can be constructed via multiple application of the triangle inequality \cite{distance1}, e.g., $d(A,B) \leq d(A,B') + d(B,B')$ and $d(B,B') \leq d(A',B') + d(A',B)$ give $d(A,B) \leq d(A,B') + d(A',B') + d(A',B)$. Here, $\{A,A'\}$ and $\{B,B'\}$ label different measurable properties of Alice's and Bob's systems, respectively,  and $d(A,B)$, $d(A,B')$, $\dots$ are information-theoretic distances between them. The natural intuition behind this inequality is that the shortest path goes directly from $A$ to $B$ and if one chose to go around through $B'$ and $A'$ one would have to take a longer route. The violation of such inequalities indicates that the system does not obey some properties of a metric and it was shown that this can happen for quantum systems \cite{distance1,distance2}. 

The notion of a distance is designed as a property between two points. In the information-theoretic framework a distance describes a relation between two random variables -- in our case, two jointly measurable observables. It is therefore somehow unnatural to expect that the same framework can be adopted for three or more jointly measurable properties. However, here we show that this can be done. 

The rest of the manuscript is organised as follows. First, we introduce an information-theoretic distance-like property that applies to more than two measurements. Next, we use this property to derive a tripartite information-theoretic Bell inequality. Then, we show that this inequality can be maximally violated within quantum theory and interpret it as an entropic version of the GHZ paradox. Finally, we discuss our results and suggest further avenues of research on this topic.

\emph{Distance-like property.} Let us consider the following function defined for binary observables $A$ and $B$
\begin{equation}
d\left(A,B\right)=H\left(A \cdot B\right),\label{eq:Distance}
\end{equation}
where $H\left(.\right)$ is the Shannon entropy. Before we proceed, we should emphasis that this function is a distance only if the outcomes of $A$ and $B$ are $\pm 1$. The measurement of $A\cdot B$ is one where the outcomes are the product of the outcomes of
$A$ and $B$, i.e. if the outcome of $A$ is $a$ and the outcome
of $B$ is $b$ , then the outcome of $A\cdot B$ is $ab$. 

The function in Eq.(\ref{eq:Distance}) satisfies all of the following
properties. 
\begin{enumerate}
\item $d\left(A,A\right)=H(A\cdot A)=0$ because the outcome of this measurement
is always one and $d\left(A,B\right)\geq0$ because $H\left(X\right)\geq0,\,\forall X$.
\item $d\left(A,B\right)=d\left(B,A\right)$. 
\item The triangle inequality: $H\left(A\cdot B\right)\leq H\left(B \cdot C\right)+H\left(A \cdot C\right),\,\forall A,B,C.$
\end{enumerate}
The triangle inequality is satisfied because $H\left(A\cdot B|A\cdot C,B\cdot C\right) = 0$, i.e. if the outcomes of the two measurements $A\cdot C$ and $B\cdot C$ are known,
then the outcome of $A\cdot B$ is the product of the two outcomes and is therefore known. More precisely, $H\left(A\cdot B\right)\leq H\left(A\cdot B,B \cdot C,A\cdot C\right)=H\left(A\cdot B|A\cdot C,B\cdot C\right)+H\left(B\cdot C,A\cdot C\right)=H\left(B\cdot C,A\cdot C\right)\leq H\left(B\cdot C\right)+H\left(A\cdot C\right).$ In the above we used the facts that $H(AB)=H(A|B) + H(B)$, $H(AB) \leq H(A) + H(B)$ and $H(A) \leq H(AB)$.

The distance in Eq. (\ref{eq:Distance}) can simply be extended to a distance-like property for
the multipartite measurements. Note, that for
a set of operators $\left\{ A_{1},A_{2},\dots,A_{n}\right\} $ one can define 
\begin{equation}\label{d}
\delta \left(A_{1},A_{2},\dots,A_{n}\right)=H\left(A_{1}\cdot A_{2} \cdot \ldots \cdot A_{n}\right),
\end{equation}
which is the natural extension of the distance for two operators. 

\emph{Tripartite information-theoretic Bell inequality.} Let us examine the properties of (\ref{d}) in the context of multipartite measurements. For multipartite measurements, $A_{i}$ would be the local observable. Within quantum theory this measurement would be represented by the local $A_{i}$
operator for the specific party conjugated with the identity operators
for other parties. For instance, for the measurement
of $A$ for Alice and \textbf{$B$} for Bob and $C$ for Charlie one would
have $A_{1}=A\otimes\mathbb{I}\otimes\mathbb{I}$, $A_{2}=\mathbb{I}\otimes B\otimes\mathbb{I}$, $A_{3}=\mathbb{I}\otimes\mathbb{I}\otimes C$
leading to $\delta \left(A_{1},A_{2},A_{3}\right)=H\left(A \otimes B \otimes C\right)$. 

The function $\delta$ is obviously symmetric, but it also has a nice associative property which is 
\begin{equation}
\delta \left(A_{1},A_{2},A_{3}\right)=d\left(A_{1},\left(A_{2}\cdot A_{3}\right)\right).\label{eq:AssociativeProperty}
\end{equation}
This comes from the fact that $H\left(A_{1}\cdot A_{2}\cdot A_{3}\right)=H\left(A_{1}\cdot \left(A_{2}\cdot A_{3}\right)\right)$.
Note that using to the symmetry property, any two $A_{i}$ could be
associated.

Now we derive the following inequality: 
\begin{eqnarray}
& &\delta \left(A_{1},B_{1},C_{1}\right)\leq  \label{eq:EntropicMermin} \\ & &\delta \left(A_{1},B_{2},C_{2}\right)+\delta \left(A_{2},B_{2},C_{1}\right)+\delta \left(A_{2},B_{1},C_{2}\right).\nonumber
\end{eqnarray}
The derivation is as follows:
\begin{align*}
\delta  & \left(A_{1},B_{1},C_{1}\right)\leq d\left(A_{1},\left(B_{2}.C_{2}\right)\right)+d\left(\left(B_{2}.C_{2}\right),\left(B_{1}.C_{1}\right)\right)\\
= & ~ d\left(A_{1},B_{2}.C_{2}\right)+\delta \left(B_{2},C_{1},B_{1},C_{2}\right)\\
\leq & ~ \delta \left(A_{1},B_{2},C_{2}\right)+d\left(A_{2},B_{2}\cdot C_{1}\right)+d\left(A_{2},B_{1}\cdot C_{2}\right)\\
= & ~ \delta \left(A_{1},B_{2},C_{2}\right)+\delta \left(A_{2},B_{2},C_{1}\right)+\delta \left(A_{2},B_{1},C_{2}\right).
\end{align*}
The first and the last inequalities come from using the triangle inequality
with $B_{2}\cdot C_{2}$ and $A_{2}$ respectively and equality in the middle
comes from the symmetry property. 

Note that the derivation holds not only for the function $\delta$, but for any distance with the associativity property in
(\ref{eq:AssociativeProperty}). For instance, applying the generalisation of the co-variance distance \cite{distance1,distance2}, $\delta\left(A_{1},A_{2},A_{3}\right)=1-\left\langle A_{1}\cdot A_{2}\cdot A_{3}\right\rangle $ to the inequality (\ref{eq:EntropicMermin}) gives the original tripartite Mermin inequality
\cite{Mermin}. 

The inequality (\ref{eq:EntropicMermin}) was derived using the classical properties of Shannon entropy, therefore it must hold in any theory that obeys them. In particular, in local realistic theories there exists a joint probability distribution for all observables $A_1,\dots,C_2$ \cite{Fine} and as a consequence there exists a joint entropy $H(A_1\dots C_2)$ which implies the validity of (\ref{eq:EntropicMermin}). However, one may expect the violation of this inequality if a theory does not admit a joint probability distribution.

\emph{Quantum violation and the paradox.} Let us consider a three-qubit system in a GHZ state $|GHZ\rangle=\frac{1}{\sqrt{2}}(|000\rangle + |111\rangle)$ shared between Alice, Bob and Charlie. Each of them performs one of the two possible local measurements on their subsystem: $A_1,A_2,B_1,\dots$ As previously discussed, we can choose $\delta(A_i,B_j,C_k)=H(A_i \otimes B_j \otimes C_k)$ (for $i,j,k=1,2$) and plug these measurements to the inequality (\ref{eq:EntropicMermin}) to obtain 
\begin{eqnarray}
& & H\left(A_{1}\otimes B_{1} \otimes C_{1}\right) \leq H\left(A_{1}\otimes B_{2} \otimes C_{2}\right) \nonumber \\&+& H\left(A_{2}\otimes B_{1} \otimes C_{2}\right) + H\left(A_{2}\otimes B_{2} \otimes C_{1}\right). \label{quantum}
\end{eqnarray}

Note that $A_i \otimes B_j \otimes C_k$ are binary $\pm 1$ observables, therefore $H(A_i \otimes B_j \otimes C_k)$ cannot exceed one. If this entropy is equal to one, then we have no knowledge of  $A_i \otimes B_j \otimes C_k$ and if it is zero, we are certain about the value of this observable. To simplify the notation we set $A \equiv A_{1}\otimes B_{2} \otimes C_{2}$, $B \equiv A_{2}\otimes B_{1} \otimes C_{2}$, $C \equiv A_{2}\otimes B_{2} \otimes C_{1}$ and $D \equiv A_{1}\otimes B_{1} \otimes C_{1}$. The inequality (\ref{quantum}) takes form
\begin{equation}\label{GHZ}
H(D) \leq H(A) + H(B) + H(C).
\end{equation}
It bounds the entropy of $D$ via the entropies of $A$, $B$ and $C$. In particular, it predicts that if $A$, $B$ and $C$ are known (their entropy is zero), then $D$ must be known too. An additional argument for that is along the GHZ reasoning --- if the outcomes are predetermined, then the values of $A$, $B$, $C$ and $D$ (denoted as $a$, $b$, $c$ and $d$) must multiply to one ($abcd=1$). Therefore, if we know $A$, $B$ and $C$, we automatically know $D$, since $d=abc$.

However, quantum theory allows for a violation of the inequality (\ref{GHZ}). If Alice, Bob and Charlie chose
\begin{eqnarray}
A_{1}=B_{1}=C_{1} & = & \cos\left(\frac{\pi}{6}\right)X+\sin\left(\frac{\pi}{6}\right)Y,\nonumber \\
A_{2}=B_{2}=C_{2} & = & \cos\left(\frac{\pi}{12}\right)X-\sin\left(\frac{\pi}{12}\right)Y,\label{eq:MerminMeasurements}
\end{eqnarray}
they would observe that $H(A)=H(B)=H(C)=0$, but at the same time $H(D)=1$. We achieved maximal algebraic violation and observed that although $A$, $B$ and $C$ are known, $D$ is completely unknown!

\emph{Test via compression.} The plausible feature of the original GHZ paradox is that it does not involve probabilities, despite the fact that quantum theory is fundamentally probabilistic. Here, the entropies $H(A)$, $H(B)$ and $H(C)$ are zero, therefore the corresponding measurement events are fully predetermined. On the other hand, $H(D)=1$ indicates that $D$ is maximally random. Still the entropic GHZ paradox can be investigated without invoking quantum probabilities by using the data compression approach proposed in \cite{Compression}.

Imagine that Alice, Bob and Charlie perform $n$ rounds of measurements $A_i$, $B_j$ and $C_k$, respectively. They produce bit strings $a^{(i)}=a_1^{(i)}a_2^{(i)}\dots a_n^{(i)}$, $b^{(j)}=b_1^{(j)}b_2^{(j)}\dots b_n^{(j)}$, and $c^{(k)}=c_1^{(k)}c_2^{(k)}\dots c_n^{(k)}$. Due to the fact that each measurement round is performed on a different independent triple of qubits prepared in the GHZ state, the subsequent bits in each string are independent and identically distributed (i.i.d.). In this case, the expression $n H(A_i\otimes B_j \otimes C_k)$ is the Shannon entropy of a bit string that is a concatenation (XOR) of bit strings $a^{(i)}$, $b^{(j)}$ and $c^{(k)}$, i.e. $H(a^{(i)} \oplus b^{(j)} \oplus c^{(k)})$.

For i.i.d. bit strings Shannon entropy gives the best possible compression rate \cite{TC}. In case of real life compressors, like gzip or Huffman code, the compression rate is worse than the Shannon entropy. Still, for uniform (deterministic) bit strings $a^{(i)} \oplus b^{(j)} \oplus c^{(k)}$ the compression rate $C(a^{(i)} \oplus b^{(j)} \oplus c^{(k)})$ obtained by a real life compressor $C$ is of the order $O(\log n)$ \cite{Bennett, CV}. It is therefore justified to use a modified version of the inequality (\ref{quantum}) 
\begin{eqnarray}
& & C(a^{(1)} \oplus b^{(1)} \oplus c^{(1)}) \leq C(a^{(1)} \oplus b^{(2)} \oplus c^{(2)}) \nonumber \\&+& C(a^{(2)} \oplus b^{(1)} \oplus c^{(2)}) + C(a^{(2)} \oplus b^{(2)} \oplus c^{(1)}) \label{compression}
\end{eqnarray}
as a valid bound on classical theories, if the compression rates on the right hand side are $O(\log n)$. The above inequality will be violated by the quantum measurements discussed in the previous section, because the bit strings on the right hand side are predicted to be uniform, whereas the bit string on the left hand side is predicted to be maximally random, therefore $C(a^{(1)} \oplus b^{(1)} \oplus c^{(1)}) = O(n)$. Although we do not provide the proof, we speculate that the above inequality will be also valid (as a classical bound) in case of experimental noise and for an arbitrary measurement scenario. 

Interestingly, the tripartite inequality (\ref{quantum}) is more robust to noise than the bipartite information-theoretic inequality studied in \cite{BC}. In case of the white noise admixture $\openone/8$ to the pure GHZ state $\rho(p)=\left(1-p\right)|GHZ\rangle\langle GHZ|+p \openone/8$,
the value of $p$ for which the violation vanishes is $p \approx 0.123$. The corresponding threshold value for the bipartite inequality \cite{BC} is $p \approx 0.04$. Of course, this comparison can be only used as a reference, since in reality it is much harder to engineer the tripartite GHZ state than the bipartite singlet state. 

\emph{Conclusions.} We proposed a tripartite information-theoretic Bell inequality based on a distance-like property. The inequality is maximally violated by measurements on three qubits in the GHZ state and we used this fact to formulate the entropic version of the GHZ paradox. Finally, we discussed the test of this paradox in terms of compressibility of bit strings generated from the measurement data obtained by parties sharing the three qubits and showed that our multipartite scenario is more robust to noise than the corresponding bipartite scenario. 

There are several open problems that require further investigation. First of all, it is natural to look for an extension of our result to more than three parties and to higher-level systems. Moreover, bipartite information-theoretic Bell inequalities are less efficient in detection of the lack of local realism than the correlation based inequalities.  It is therefore important to investigate the class of states violating our inequality and to compare this class with the class of states violating the Mermin inequality. Finally, it would be important to prove (or disprove) our conjecture that the inequality (\ref{compression}) is always valid.

\emph{Acknowledgements.} S.R. was supported by Canada’s NSERC, 
MPrime, CIFAR, and CFI and IQC. 
P. K. and D. K. are supported by the Foundational Questions Institute (FQXi) and by the National Research Foundation and Ministry of Education in Singapore.


\begin{thebibliography}{99}

\bibitem{Bell} J. S. Bell, Physics (N.Y.) {\bf 1}, 195 (1965).

\bibitem{Mermin} N. D. Mermin, Phys. Rev. Lett. {\bf 65}, 1838 (1990).

\bibitem{M2} M. Ardehali, Phys. Rev. A {\bf 46}, 5375 (1992). 

\bibitem{M3} A. V. Belinskii and D. N. Klyshko, Phys. Usp. {\bf 36}, 653 (1993).

\bibitem{GHZ} D. M. Greenberger, M. A. Horne, and A. Zeilinger, {\it Bell's Theorem, Quantum Theory, and Conceptions of the Universe}, M. Kafatos (Ed.), Kluwer, Dordrecht, 69-72 (1989).

\bibitem{BC} S. L. Braunstein and C. M. Caves, Phys. Rev. Lett. {\bf 61}, 662(1988).

\bibitem{distance1} P. Kurzy\'nski and D. Kaszlikowski, Phys. Rev. A {\bf 89}, 012103 (2014).

\bibitem{distance2} B. W. Schumacher, Phys. Rev. A {\bf} 44, 7047 (1991).

\bibitem{Zurek} W. H. Zurek, Nature {\bf 341}, 119 (1989).

\bibitem{Fine} A. Fine, Phys. Rev. Lett. {\bf 48}, 291 (1982).

\bibitem{Compression} P. Kurzynski, M. Markiewicz, and D. Kaszlikowski, arXiv:1310.5644 (2013).

\bibitem{TC} T. M. Cover and J. A. Thomas, {\it Elements of information theory}, Wiley (1991).

\bibitem{Bennett} C. H. Bennett, P. Gacs, M. Li, P. M. B. Vitanyi, and W. Zurek, IEEE Transactions on Information Theory {\bf 44}, 1407 (1998).

\bibitem{CV} R. Cilibrasi and P. M. B. Vitanyi, IEEE Transactions on Information Theory {\bf 51}, 1523 (2005).





\end{thebibliography}
\end{document}